\documentstyle[12pt]{article}

\begin{document}
\date{}

\title{A mechanism for dynamical generation of
SU(2) Georgi-Glashow model}

\author{V.E. Rochev\\
{\it Institute for High Energy Physics,}\\{\it Protvino,
Moscow region, Russia}}

\maketitle

\newcommand{\be}{\begin{equation}}
\newcommand{\ee}{\end{equation}}
\newcommand{\ba}{\begin{eqnarray}}
\newcommand{\ea}{\end{eqnarray}}

\begin{abstract}
A mechanism for the dynamical mass generation of a 
non-Abelian gauge field  
which is based on taking into account 
the contributions of the gauge field vacuum configurations 
into the formation of the physical vacuum is considered. 
For a model of the physical vacuum as a superposition of 
Abelian configurations the gauge field propagator is 
calculated in the leading order of $1/d$-expansion 
($d$ is a space-time dimension).
One-particle spectrum of the model corresponds to 
the gauge sector of $SU(2)$ Georgi-Glashow model.
\end {abstract}

\section{Introduction}
In the Standard Model the central role belongs to the Higgs 
mechanism which gives the masses to particles without 
violating the cardinal principles of theories, such as 
local gauge invariance and renormalizability. The Higgs 
mechanism  is a well adjusted and efficient machine,
and its description rightfully occupies its honorable place 
in any textbook on the modern  high energy physics.
However, in spite of all experimental efforts, no traces of 
the scalar sector of the Standard Model have been found 
hitherto, and the problem on searching alternatives to 
the Higgs mechanism  is quite actual.

The generation of mass of a gauge 
field and the spontaneous symmetry breaking connected with 
it are defined by the 
structure of the ground state of the theory, that is a 
physical vacuum. Modeling of this structure is the main 
problem of the dynamical symmetry breaking description  
(see, for instance, \cite{Mir}).

In this paper a mechanism for the dynamical mass generation 
of the non-Abelian gauge field is considered which does not 
require entering a scalar field and  other additional 
fields. Here the gauge field itself, more exactly its vacuum 
constituent undertakes the role of an order parameter, which 
is played by the vacuum expectation of a scalar field in 
the usual Higgs mechanism. Appearance of such a vacuum 
constituent is a manifestation of a nontrivial structure 
of physical vacuum of the quantum field theory.
This vacuum constituent arises quite naturally in 
constructing an iteration solution of the Schwinger-Dyson 
equations by the method, suggested in \cite{Ro1}, 
\cite{Ro2}, which we shall use in the present work as well.

To illustrate the method we examine 
the spontaneous symmetry breaking in the scalar theory. 
(As is well known, this phenomenon is a foundation for the 
Higgs mechanism  in the Standard Model.)
Consider the theory of a scalar field $\phi$ with the 
Lagrangian
\be
{\cal L} = \partial_\mu \phi^* \partial_\mu \phi -
m^2 \phi^* \phi - \frac{\lambda}{2} (\phi^* \phi)^2.
\label{Lagscal}
\ee
(We are working in the Minkowsky metric with 
$g_{00}=1$, but will not distinguish between
the upper and lower
indices to simplify the notations.)

The Schwinger-Dyson equation
(see, for instance,
 \cite{Rob})
 for the generating functional 
$G(j)$ of the Green functions
of this model has the form
\be
\frac{\lambda}{i}\frac{\delta^3 G}
{\delta j^* \delta j \delta j^*}
- (m^2 + \partial^2) \frac{1}{i} \frac{\delta G}{\delta j^*}
+ j G = 0.\label{SDEscal}
\ee
Here $j(x)$ is a source of the field $\phi^* (x)$. 
At $\lambda = 0$ eq.(\ref{SDEscal}) has the
unique (up to a normalization factor)  solution
$$
G^{pert} = \exp \{ i \int dx dy j^*(x) \Delta_c(x-y) j(y)\},
$$
where $\Delta_c = (m^2 + \partial^2)^{-1}$ is the free 
propagator.

This solution is a foundation for the iterative solution 
of eq.(\ref{SDEscal}) in the form of the perturbation series.
At $m^2<0$ such a solution is unstable --- the physical 
vacuum in this case differs from the trivial vacuum of 
the perturbation theory. To describe the solution in this 
case, which corresponds the spontaneous symmetry breaking, 
let us consider another iterative scheme: an expansion 
near the point $j=0$.
As a leading approximation, consider the equation
\be
\frac{\lambda}{i}\frac{\delta^3 G^{(0)}}{\delta j^* \delta j 
\delta j^*}
- (m^2 + \partial^2) \frac{1}{i} \frac{\delta G^{(0)}}
{\delta j^*} = 0,
\label{LAscal}
\ee
and the iterative expansion for the generating functional
$$
G = G^{(0)} + G^{(1)} + \cdots + G^{(n)} + \cdots,
$$
is constructed by the subsequent solution of the iteration 
scheme equations
\be
\frac{\lambda}{i}\frac{\delta^3 G^{(n)}}{\delta j^* \delta j 
\delta j^*}
- (m^2 + \partial^2) \frac{1}{i} \frac{\delta G^{(n)}}
{\delta j^*} =
- j G^{(n-1)} .\label{ISscal}
\ee
 Leading approximation equation
(\ref{LAscal}) has the solution
$$
G^{(0)} (j) = \exp i\int dx [v^*j + vj^*],
$$
where $v$ obeys the "characteristic equation"
\be
(\lambda v^* v + m^2 + \partial^2)v = 0.
\label{Charscal}
\ee
Characteristic equation (\ref{Charscal}) 
has the form of a classical field equation of the theory 
with Lagrangian (\ref{Lagscal}), and its solution
 $v$ is the vacuum expectation of the field
 $\phi$ in the leading approximation of the given iteration 
 scheme.

The trivial solution
 $v\equiv0$ corresponds to the trivial perturbative vacuum,
 and the corresponding iterative expansion is the 
 perturbation 
 series, i.e. the usual perturbation theory is a partial
 case of the given iteration scheme.
 At $m^2<0,\,\lambda>0$,  the Green functions of 
 the perturbation
 theory maintain tachyon poles, which indicates 
 the instability of the trivial vacuum, but 
 in this case  another class of constant solutions of 
 characteristic equation (\ref{Charscal}) exists
$$
v^*v = -m^2/\lambda.
$$
The iteration scheme based on this solution is a 
perturbation theory over spontaneously broken 
non-perturbative vacuum.
For the construction of the iterations in this 
case, it is convenient to go over to the new sources
$$
j_+ = \frac{1}{\sqrt{2v^*v}}(v^*j+vj^*),\;
j_- = \frac{-i}{\sqrt{2v^*v}}(v^*j-vj^*).
$$
In terms of these sources we obtain the solution 
of the first-step iteration scheme equations
as $G^{(1)}=P^{(1)}G^{(0)}$, where
$$
P^{(1)} = \frac{i}{2}\int dxdy\{j_+\Delta_Hj_+ + 
j_-\Delta_Gj_-\}.
$$
Here $\Delta_H=(2\lambda v^*v+\partial^2)^{-1}$
is the  Higgs boson propagator, and $\Delta_G=\partial^{-2}$
is the Goldstone boson propagator.

Therefore, the adequate choice of the vacuum constituent
--- the solution $v$ of characteristic equation 
(\ref{Charscal})
--- defines the adequate structure of Green 
functions and one-particle spectrum of the theory.

We shall apply this scheme to the construction 
of the iterative solution for a non-Abelian gauge theory.
Corresponding characteristic equations have a great number
of various solutions, and a choice of a class of 
the solutions, i.e. the vacuum constituents,
defines a choice of the candidate to the physical 
vacuum of the theory.
Here we consider the simplest non-trivial 
class of the solutions ("Abelian solutions", see below) 
and demonstrate 
that in the leading order of $1/d$-expansion this class 
of solutions leads to the dynamical realization of 
$SU(2)$ Georgi-Glashow model \cite{GG}.

\section{Iteration scheme for non-Abelian gauge theory}

Consider the theory of a gauge field
${\bf A}_\mu\equiv~A^a_\mu$ with the Lagrangian
\be
{\cal L} = -\frac{1}{4}{\bf F}_{\mu\nu}
{\bf F}_{\mu\nu}({\bf A})
- \frac{1}{2\alpha}(\partial_\mu{\bf A}_\mu)^2
- {\bf\bar c}\partial_\mu{\cal D}_\mu({\bf A}){\bf c}.
\label{Lagr}
\ee
Here ${\bf F}_{\mu\nu}({\bf A})\equiv~F^a_{\mu\nu}({\bf A})
=\partial_{\mu}A^a_{\nu}-\partial_{\nu}A^a_{\mu}
+gf^{abc}A^b_{\mu}A^c_\nu$
is the gauge field tensor;
${\cal D}_\mu({\bf A})\equiv{\cal D}^{ab}_{\mu}({\bf A})
=\delta^{ab}\partial_{\mu}-
gf^{abc}A^c_{\mu}$ 
is the covariant derivative;
$\alpha$ 
is a gauge parameter;
 ${\bf c}\equiv~c^a$ 
 is the Faddeev-Popov ghost field;
 $f^{abc}$ 
 are structure constants of the gauge group.

Let us introduce the generating functional $G(J)$ of 
Green functions which depends on the gauge field source
  ${\bf J}_\mu\equiv~J^a_\mu$ 
and the ghost field sources  
 ${\bf j}\equiv~j^a$ and ${\bf\bar j}\equiv~{\bar j}^a$.
The generating functional $G$
is a solution of the system of equations in functional 
derivatives --- the Schwinger-Dyson equations:
\be
{\cal D}_{\nu}({\cal A})
{\bf F}_{\nu\mu}({\cal A}) G +
 \frac{1}{\alpha}\partial_{\mu}\partial_{\nu}{\cal A}_\nu G 
+ g{\bf fC}
\partial_{\mu}{\bf\bar C}G + {\bf J}_{\mu} G = 0,
\label{SDEA}
\ee
\be
\partial_{\mu} {\cal D}_{\mu}({\cal A}){\bf C} G = {\bf j} G.
\label{SDEghost}
\ee
Here the following notations are introduced
$$
{\cal A}_\mu\equiv{\cal A}^a_\mu = 
\frac{1}{i}\frac{\delta}{\delta J^a_\mu},\;
{\bf C}\equiv C^a = 
\frac{1}{i}\frac{\delta}{\delta {\bar j}^a},\;
{\bf\bar C}\equiv {\bar C}^a = i\frac{\delta}{\delta j^a}.
$$

 For solving  equations
(\ref{SDEA})-(\ref{SDEghost}) we shall use
the iteration scheme of
 \cite{Ro1}, \cite{Ro2}.
The leading approximation of this scheme is the
system of equations 
\be
{\cal D}_{\nu}({\cal A})
{\bf F}_{\nu\mu}({\cal A}) G^{(0)} +
 \frac{1}{\alpha}\partial_{\mu}\partial_{\nu}
 {\cal A}_\nu G^{(0)} 
+ g{\bf fC}\partial_{\mu}{\bf\bar C}G^{(0)} = 0,
\label{LAA}
\ee
\be
\partial_{\mu} {\cal D}_{\mu}({\cal A}){\bf C} G^{(0)} = 0.
\label{LAghost}
\ee
The equations of the $n$-th step of the iteration 
scheme have the form
\be
\biggl\{{\cal D}_{\nu}({\cal A})
{\bf F}_{\nu\mu}({\cal A})  +
 \frac{1}{\alpha}\partial_{\mu}\partial_{\nu}{\cal A}_\nu 
+ g{\bf fC}
\partial_{\mu}{\bf\bar C}\biggr\}G^{(n)} = -{\bf J}_{\mu} 
G^{(n-1)},
\label{ISA}
\ee
\be
\partial_{\mu} {\cal D}_{\mu}({\cal A}){\bf C} G^{(n)} =
{\bf j} G^{(n-1)}.
\label{ISghost}
\ee
The solution of leading approximation equations
(\ref{LAA})-(\ref{LAghost})
is 
\be
G^{(0)} = \exp i\Bigl\{{\bf J}_\mu\star{\bf V}_\mu + 
{\bf\bar j}\star{\cal G}
+ \bar{\cal G}\star{\bf j}\Bigr\},
\label{G0}
\ee
where 
${\bf J}_\mu\star{\bf V}_\mu\equiv\int~dx~J_\mu^a(x)
V_\mu^a(x)$,
etc.

Coefficient functions
${\bf V}_\mu\equiv~V_\mu^a$ ³ ${\cal G}\equiv{\cal G}^a$
(vacuum constituents) are solutions of the system of 
characteristic equations
\be
{\cal D}_{\nu}({\bf V})
{\bf F}_{\nu\mu}({\bf V})  +
 \frac{1}{\alpha}\partial_{\mu}\partial_{\nu}{\bf V}_\nu 
+ g{\bf f}{\cal G}
\partial_{\mu}\bar{\cal G} = 0,
\label{CharA}
\ee
\be
\partial_{\mu} {\cal D}_{\mu}({\bf V}){\cal G} = 0.
\label{Charghost}
\ee
The solution of the $n$-th-step equations of the 
iteration scheme
has the form
$$
G^{(n)} = P^{(n)} G^{(0)},
$$
and  taking into account characteristic equations
(\ref{CharA})-(\ref{Charghost}), 
we obtain the system of equations for the functional
 $P^{(n)}$
\ba
\biggl\{\Bigl({\cal D}_{\nu}({\bf V}) - 
g{\bf f}{\cal A}_\nu\Bigr)
\Bigl({\cal D}_\nu({\bf V})
{\cal A}_\mu-{\cal D}_\mu({\bf V}){\cal A}_\nu
+ g{\bf f}{\cal A}_\mu{\cal A}_\nu\Bigr)
-g{\bf f}{\bf F}_{\mu\nu}({\bf V})
{\cal A}_\nu \nonumber\\ +
 \frac{1}{\alpha}\partial_{\mu}\partial_{\nu}{\cal A}_\nu 
+ g{\bf f}\Bigl({\bf C}\partial_{\mu}{\bf\bar C} +
{\cal G}\partial_{\mu}{\bf\bar C} + 
{\bf C}\partial_{\mu}\bar{\cal G}
\Bigr)\biggr\}P^{(n)}
=  -{\bf J}_{\mu} P^{(n-1)},\nonumber\\
\partial_{\mu}\biggl\{\Bigl({\cal D}_{\mu}({\bf V}) 
- g{\bf f}{\cal A}_\mu\Bigr)
{\bf C} - g{\bf f}{\cal A}_\mu{\cal G}\biggr\}P^{(n)} = 
{\bf j} P^{(n-1)}.\nonumber
\ea
Since $P_0 = 1$,
the solution of this system for any $n$ is a polynomial in 
sources ${\bf J}_\mu$ and ${\bf j}$.
Coefficient functions of this polynomial define the Green 
functions. 
 At each step the  equations of the iteration scheme give 
 a closed  system of equations for these  functions. 
 The solution of the first-step equations  is a 
 quadratic polynomial, defining two-point functions 
 (the propagators). At the second step the solution 
 is a polynomial of the fourth degree, defining 
 three-and four-point functions, as well as corrections 
 to the propagators, etc.
To eliminate ultraviolet divergences, it is necessary 
to modify the equations of the iteration scheme by 
introducing the corresponding counterterms 
(see \cite{Ro1}). Note that functions of the 
leading approximation and of the
first step are  ultraviolet-finite.

\section{Physical vacuum and vacuum constituents}

In the leading approximation of considered iteration 
scheme
the ground state (the physical vacuum of the theory) 
is defined by a choice  of solutions of the 
characteristic equation system
(\ref{CharA})-(\ref{Charghost}).
Each solution
  $V\equiv({\bf V}_\mu,{\cal G})$ of the characteristic 
  equation system defines a partial solution
  $G_V(J)$ of the iteration scheme.
This solution will be referred to as  corresponding 
to a partial mode
$\mid V>$.

The trivial solution 
 ${\bf V}_\mu={\cal G}=~0$
corresponds to the leading approximation
 $G^{(0)}=1$.
The iteration scheme based on this solution is a 
perturbation theory in the coupling  constant over 
the trivial perturbative vacuum. Nontrivial solutions of  
the characteristic equations define non-perturbative modes.
These solutions have a sense of the vacuum constituents of
the quantum fields ${\bf A}_\mu$ and ${\bf c}$, like the 
vacuum constituent $v$ of  scalar field in the Higgs 
mechanism ( see Introduction).

The choice of the approximation for the description of 
the physical vacuum $\mid~0>$ must ensure general 
physical requirements, such as Poincar{\' e}-invariance, 
cluster decomposition, etc.
In the Higgs mechanism  it is sufficient to take  the 
constant solution $v=\sqrt{-m^2/\lambda}$ for this purpose.

In the case under consideration the situation is more 
complicated. Obviously, the choice of a separate 
partial mode with ${\bf V}_\mu\neq~0$ as a 
leading approximation to the physical vacuum 
("a candidate for the physical vacuum") does not 
ensure Poincar{\' e}-invariance of the theory. Notice, 
however, that Schwinger-Dyson equations 
(\ref{SDEA})-(\ref{SDEghost}) are the linear 
functional-differential equations for the 
generating functional, and any superposition 
of partial solutions $\sum G_V(J)$ 
is also a solution of these equations. So we can 
choose a superposition of partial modes as a 
candidate for the physical vacuum, and choose 
the generating functional of the physical Green 
functions as the superposition  
$$
<0\mid 0>_J = G(J) = \sum_{\{V\}} G_V(J),
$$
corresponding to a class 
 $\{V\}$ of solutions of the characteristic 
 equations.
We shall suppose  this superposition can be chosen in 
such a way 
that all the contributions, breaking the 
Poincar\'e-invariance, 
are mutually
canceled, and the resulting  theory turns out to be 
Poincar{\' e}-invariant. 
For instance, the  expectation value of the gauge field 
must disappear
 $$
<0\mid {\bf A}_\mu\mid 0> = \frac{1}{i}\frac{\delta G}
{\delta {\bf J}_\mu}
\Bigg\vert_{J=0} = 0,
$$
in spite of  the contributions of separate partial modes 
in this vacuum expectation can be different from  zero.
Further, the higher derivatives of the  physical generating 
functional, defining multipoint functions, must be  
translational-invariant after switching off the sources, etc.
The set of these conditions will ensure the
Poincar\'e-invariance of the theory.

Consider  energetics.
For  energy density  $E$ of  ground state exists the 
well-known formula

\be
E = i\log G\bigg/\int dx.
\label{E}
\ee
This formula defines, in essence, a change of 
normalization for  generating functional, and its direct
application
is difficult.
A normalization-independent formula for the energy 
can be obtained considering $E$ as a function of
parameters of Lagrangian. Differentiating 
eq.(\ref{E}) over $g$ we get the formula
\be
\frac{\partial E}{\partial g} = \frac{1}{G}\int dx
\biggl\{\frac{1}{2}F_{\mu\nu}^a({\cal A})
f^{abc}{\cal A}_\mu^b
{\cal A}_\nu^c + 
{\bar C}^af^{abc}\partial_\mu{\cal A}_\mu^bC^c\biggr\}
G\Bigg/\int dx,
\label{dE/dg}
\ee
which is normalization-independent. 
For the leading approximation of our 
iteration scheme we obtain
\be
\frac{\partial E_0}{\partial g} = \int dx
\biggl\{\frac{1}{2}F_{\mu\nu}^a({\bf V})f^{abc}V_\mu^b
V_\nu^c + 
\bar{\cal G}^af^{abc}\partial_\mu V_\mu^b{\cal G}^c\biggr\}
\Bigg/\int dx.
\label{dE0/dg}
\ee
for the energy of  partial mode
$\mid V>$.
Obviously  a variety of partial modes, 
defining some candidate for  physical vacuum,
must possess the same energy.

\section{Ward-Slavnov-Taylor identities}

Gauge invariance imposes  the restrictions on Green 
functions which are known as Ward-Slavnov-Taylor
identities. From the Jacobi identities for the structure 
constants of the gauge group  the identity follows
\be
{\cal D}_\mu({\cal A}){\cal D}_\nu({\cal A})
{\bf F}_{\mu\nu}({\cal A}) \equiv 0.
\label{WI}
\ee
Acting by operator ${\cal D}_\mu({\cal A})$ 
on Schwinger-Dyson equation (\ref{SDEA}) and taking 
into account (\ref{WI}) we obtain
 the generating relation for the Ward-Slavnov-Taylor 
 identities 
\be
 \frac{1}{\alpha}{\cal D}_\mu({\cal A})
\partial_{\mu}\partial_{\nu}{\cal A}_\nu G 
+ g{\cal D}_\mu({\cal A}){\bf fC}
\partial_{\mu}{\bf\bar C}G = -{\cal D}_\mu({\cal A})
{\bf J}_{\mu} G.
\label{WIGR}
\ee
Differentiating this relation and switching off the 
sources, we get the desired restrictions on the Green 
functions. As  relation (\ref{WIGR}) is an identity, 
and $G_V(J)$ is a solution of 
Schwinger-Dyson equations, these restrictions must 
be fulfilled for each separate partial mode.

If ${\bf V}_\mu\neq~0$  or ${\cal G}\neq~0$ their 
form is certainly different from that of the usual
Ward-Slavnov-Taylor identities. Relation (\ref{WIGR}) 
must be fulfilled in each order
of the considered
iteration scheme that is the chain of relations of 
type (\ref{WIGR}) must be satisfied, where in the 
left hand side $G$ is changed for $G^{(n)}$, while in 
the right hand side -- for $G^{(n-1)}$. In the 
leading approximation the consequence of  
(\ref{WIGR}) is the condition
\be
 \frac{1}{\alpha}{\cal D}_\mu({\bf V})
\partial_{\mu}\partial_{\nu}{\bf V}_\nu 
+ g{\cal D}_\mu({\bf V}){\bf f}{\cal G}
\partial_{\mu}\bar{\cal G} = 0
\label{WIChar}
\ee
on the solutions ${\bf V}_\mu$ and ${\cal G}$
of characteristic equations (\ref{CharA})-(\ref{Charghost}).

\section{Abelian configurations}

Characteristic equations (\ref{CharA})-
(\ref{Charghost}) possess a rich
ensemble of various solutions, that is a
reflection of the complex vacuum structure of 
non-Abelian gauge theory. In this paper we restrict 
ourselves  to the analysis of the simplest solutions 
of these equations, bringing, however, to nontrivial
physical effects.
First of all note  that if we  impose the 
subsidiary condition
\be
\partial_\mu{\bf V}_\mu = 0,
\label{Subcond}
\ee
then, as is seen from (\ref{WIChar}), for the 
ghost vacuum field  ${\cal G}$ we can restrict
ourselves to the trivial solutions
${\cal G}=\bar{\cal G}=0$ without contradicting  
the gauge invariance condition.
Further, there exist a class of the simplest 
solutions of characteristic equation (\ref{CharA}),
namely,
"Abelian" solutions, for which  the dependencies on the 
space-time coordinates and on isotopic variables are 
separated:
\be
{\bf V}_\mu(x) = {\bf n}V_\mu(x),
\label{AS}
\ee
where ${\bf n}$ is a unit vector in the isotopic space.
For such vacuum configurations all nonlinear 
(non-Abelian) terms in eq. 
(\ref{CharA})
disappear, and, with   condition
(\ref{Subcond}) taken into account, this equation 
becomes  the d'Alambert equation 
$
\partial^2V_\mu~=~0.
$

For Abelian solutions (\ref{AS}) from formula 
(\ref{dE0/dg}) we get
$$\frac{\partial E_0}{\partial g} = 0,$$
then for this class of solutions the ground state 
energy  in leading approximation does not depend on 
coupling constant. In this sense such class of solutions
can be named "quasiperturbative".

\section{First-step equations}

Equations of the first step define gauge field and 
ghost field propagators. Polynomial $P^{(1)}$ is 
quadratic in  sources and at ${\cal G}=\bar{\cal G}=0$ 
has  the form 
\be
P^{(1)} = \frac{1}{2i}{\bf J}_\mu\star{\bf D}_{\mu\nu}
\star{\bf J}_\nu
+ i{\bf\bar j}\star{\bf D}\star{\bf j}.
\label{P1}
\ee
Here ${\bf D}_{\mu\nu}\equiv~D_{\mu\nu}^{ab}
(x,y\mid{\bf V})$ is the gauge field propagator; 
${\bf D}\equiv~D^{ab}(x,y\mid{\bf V})$ --
the ghost propagator.

The iteration scheme gives the following equation for 
${\bf D}$
\be
\partial_\mu{\cal D}_\mu({\bf V}){\bf D} = 1.
\label{eqD}
\ee
From generating relation (\ref{WIGR}) one obtains the 
following relation for the longitudinal part of 
${\bf D}_{\mu\nu}$ 
\be
\frac{1}{\alpha}{\cal D}_\mu({\bf V})
\partial_\mu\partial_\nu{\bf D}_{\nu\lambda}
= {\cal D}_\lambda({\bf V}).
\label{WID}
\ee
Since from subsidiary condition (\ref{Subcond}) it 
follows that
$[\partial_\mu,{\cal D}_\mu({\bf V})] = 0$, then 
from (\ref{eqD}) and (\ref{WID}) we get
\be
\partial_\nu{\bf D}_{\nu\lambda} 
= \alpha{\bf D}\star{\cal D}_\lambda({\bf V}).
\label{long}
\ee
At ${\bf V}_\mu=0:\;\;{\bf D}=\partial^{-2}$, and 
identity (\ref{long})
gains the familiar form
$$
\partial_\nu{\bf D}_{\nu\lambda} 
= \alpha\frac{\partial_\lambda}{\partial^2}.
$$
With taking into account (\ref{long}), from the 
iteration scheme equations we obtain the equation 
for ${\bf D}_{\mu\nu}$:
\be
{\cal P}_{\mu\nu}({\bf V}){\bf D}_{\nu\lambda} =
g_{\mu\lambda} -
 \partial_\mu{\bf D}\star{\cal D}_\lambda({\bf V}),
\label{eqDmunu}
\ee
where
$
{\cal P}_{\mu\nu}({\bf V}) \equiv
\biggl\{{\cal D}^2({\bf V})g_{\mu\nu} - 
{\cal D}_\mu({\bf V}){\cal D}_\nu({\bf V})
+ 2\Bigl[{\cal D}_\mu({\bf V}),
{\cal D}_\nu({\bf V})\Bigr]\biggr\}.
$

As has been pointed above, the four-point and 
three-point functions enter  the 
second-step equations of the iteration scheme.
 So, for the four-point functions 
 ${\cal F}_{\mu\nu\sigma\rho}^{abcd}(x,y,z,t)$
 of the gauge field, we get the equation
\ba
{\cal P}_{\mu\nu}({\bf V})
{\cal F}_{\nu\lambda\sigma\rho}(x,y,z,t) =
-(\{g_{\mu\lambda}\delta(x-y)
{\bf 1}\otimes{\bf D}_{\sigma\rho}(z,t)\} + \nonumber\\
\{y\leftrightarrow z,\,
\lambda\leftrightarrow\sigma\} + 
\{y\leftrightarrow t,\,\lambda\leftrightarrow\rho\}).
\label{F}
\ea

Below we shall work in the transverse gauge 
$\partial_\nu{\bf D}_{\nu\lambda}=0$. 
Besides, we restrict ourselves to the  
consideration of Abelian configurations  
(\ref{AS}) of the gauge group $SU(2)$.
For the Abelian configurations  it is 
convenient to introduce the orthogonal basis
\be
u_0^{ab} = n^an^b,\;\;u_\pm^{ab} = 
\frac{1}{2}\bigl(\delta^{ab}
- n^an^b \pm i\epsilon^{abc}n^c\bigr).
\label{basis}
\ee
In basis (\ref{basis}) it is easy to separate the 
isotopic structure from the space-time one
$$
{\bf D}_{\mu\nu} = {\bf u}_0D_{\mu\nu}^{0}
+ {\bf u}_+D_{\mu\nu} + {\bf u}_-\bar{D}_{\mu\nu},
$$
$$
{\bf D} = {\bf u}_0D^{0}
+ {\bf u}_+D + {\bf u}_-\bar{D}.
$$

For $D^{0}_{\mu\nu}$ ³ $D^{0}$ we get the free 
propagator equations, that is
$$
D^{0}_{\mu\nu} = \frac{1}{\partial^2}\bigl(g_{\mu\nu}
- \frac{\partial_\mu\partial_\nu}{\partial^2}\bigr),\;\;
D^{0} = \frac{1}{\partial^2}.
$$
The equation for $D$ has the form
\be
\partial_\mu{\cal D}_\mu(V) D = 1.
\label{D}
\ee
Here ${\cal D}_\mu(V) = \partial_\mu + igV_\mu$ is 
the "Abelian" covariant derivative. Equation for  
$\bar{D}$ is obtained from (\ref{D})
by the substitution
${\cal D}_\mu\rightarrow{\cal D}_\mu^*,\;
D\rightarrow\bar{D}$.
 For $D_{\mu\nu}$ we get the equation:
\be
\biggl\{{\cal D}^2(V)g_{\mu\nu}-
{\cal D}_\mu(V){\cal D}_\nu(V) 
+ 2\Bigl[{\cal D}_\mu(V),
{\cal D}_\nu(V)\Bigr]\biggl\}D_{\nu\lambda}
= g_{\mu\lambda} - \partial_\mu D\star{\cal D}_\lambda(V).
\label{Dmunu}
\ee
The equation for $\bar{D}_{\mu\nu}$ is obtained from 
(\ref{Dmunu}) by the substitution 
${\cal D}_\mu\rightarrow{\cal D}_\mu^*,\;
D\rightarrow\bar{D},
\;D_{\mu\nu}\rightarrow\bar{D}_{\mu\nu}$.

\section{Vacuum of Abelian configurations and 1/d expansion}

As  has been pointed in Section 3, 
 when considering the nonperturbative modes
with ${\bf V}_\mu\neq~0$
it is necessary to take a  superposition of 
the non-perturbative modes as a candidate for the 
physical vacuum $\mid~0>$ 
in order to preserve the Poincar{\' e}-invariance.
As the simplest nontrivial variant for such a 
superposition, we will consider in the leading approximation 
a set of Abelian configurations $\{V\}$
  corresponding to  Abelian solutions (\ref{AS}):
$$
G^{(0)}(J) = \sum_{\{V\}} G^{(0)}_V(J) =
\sum_{\{{\bf V}\}}\exp~i{\bf J}_\mu\star{\bf V}_\mu.
$$
The operation $\sum_{\{V\}}$ must be chosen in 
such a manner that all the  Poincar{\' e}-non-invariant 
contributions would disappear, in particular, the 
conditions below must be fulfilled 
\be
<0\mid {\bf V}_\mu\mid 0> = \frac{1}{i}\frac{\delta G^{(0)}}
{\delta {\bf J}_\mu}
\Bigg\vert_{J=0} = 0,
\label{veV}
\ee
\be
<0\mid {\bf V}_\mu(x){\bf V}_\nu(y) \mid 0> = 
-\frac{\delta^2 G^{(0)}}{\delta{\bf J}_\nu(y)\delta 
{\bf J}_\mu(x)}
\Bigg\vert_{J=0} = {\bf n}\otimes{\bf n}\cdot 
f_{\mu\nu}(x-y).
\label{veV2}
\ee
It is not difficult to  make  condition (\ref{veV})
 true. For this we notice that for the Abelian 
configurations $-{\bf V}_\mu$ is a solution of the 
characteristic  equation as well as ${\bf V}_\mu$ is, 
so for obeying (\ref{veV}) it is sufficient to take the
superposition
$G^{(0)}_V(J)+G^{(0)}_{-V}(J)\sim 
\cos~{\bf J}_\mu\star{\bf V}_\mu$, or, in the general case,
$\sum_{\{V\}}\cos~{\bf J}_\mu\star{\bf V}_\mu$.
Note, that simultaneously the vacuum expectations of 
all the odd monomials in ${\bf V}_\mu$  also turn to zero:
$<0\mid~{{\bf V}_\mu}_1\cdots{{\bf V}_\mu}_{2n+1}\mid~0>=0$.
 Requirement (\ref{veV2}) is less trivial. 
It is clear that required operation 
$\sum_{\{{\bf V}\}}$ should be continual, i.e., should
correspond to some integration.
 But for the calculation of  the vacuum expectation 
 itself there is no necessity to specify this operation,
 if one is limited to configurations, for which
 $$
{\bf V}^2 \equiv  V_\mu^a(x) V_\mu^a(x) = {\cal V}^2 = 
const,
$$
that are the "equal-length" configurations 
\footnote{The quantity
${\cal V}^2$ 
plays a role of the order parameter, and its sign 
must be defined from physical considerations.}. 
Really, due to the characteristic equations and 
condition (\ref{Subcond}), the function 
$f_{\mu\nu}(x)$ must be a solution of the 
d'Alambert equation $\partial^2f_{\mu\nu}=0$ 
with the subsidiary condition 
$\partial_{\mu}f_{\mu\nu}=0$ and the initial condition
$f_{\mu\mu}(0)={\cal V}^2$.
The solution is unique 
\be
f_{\mu\nu} = \frac{{\cal V}^2}{d}g_{\mu\nu}.
\label{fmunu}
\ee
Similarly, for the four-point monomial we get:
\be
<0\mid V_\mu(x)V_\nu(y)V_\rho(z)V_\sigma(t)\mid 0> =
\frac{({\cal V}^2)^2}{d(d+2)}\bigl(g_{\mu\nu}g_{\rho\sigma}
+
g_{\mu\rho}g_{\nu\sigma} + g_{\mu\sigma}g_{\nu\rho}\bigr).
\label{veV4}
\ee

Let us turn to propagators. It is clear that the solutions 
of equations (\ref{D}) and (\ref{Dmunu}) can not be 
interpreted as particle propagators in the
Poincar{\' e}-invariant theory. Physical  
propagators must be built by means of the same 
operations of partial mode superposition:
$
D(x-y\mid{\cal V}^2) = \sum_{\{V\}} 
D(x,y\mid{\bf V}).
$

Full solving of eqs. (\ref{D}) and (\ref{Dmunu}) 
with consequent transition to the physical vacuum 
presents  a difficult problem. For its approximate 
solving  notice that, as can be seen from formulae 
(\ref{veV2}), (\ref{fmunu}) and (\ref{veV4}),  in the 
vacuum of Abelian configurations a small parameter 
arises, namely, $1/d$, where $d$ is the dimension of 
space-time. It is easy to see that if  one takes 
$D_0=\partial^{-2}$ as a leading approximation 
for  equation (\ref{D}), then after turning to the 
physical vacuum all the subsequent terms in the iterative 
solution $D=D_0+D_1+\cdots$ have a higher order 
in the parameter $1/d$. Slightly  bulkier, but  
not complicated calculation shows that for $D_{\mu\nu}$ 
(see eq.(\ref{Dmunu})) the leading approximation of 
$1/d$-expansion is 
$(\partial^2-g^2{\cal V}^2)^{-1}\star\pi_{\mu\nu}$ 
(here $\pi_{\mu\nu}$ is a transverse projector).

Thus, in the leading order in
 $1/d$, we get in the momentum space
\be
D(p) = {\bar D}(p) = -\frac{1}{p^2} + {\cal O}(1/d),
\label{ghost}
\ee    
\be
D_{\mu\nu}(p)={\bar D}_{\mu\nu}(p) = 
-\frac{1}{p^2+g^2{\cal V}^2}\Bigl(g_{\mu\nu} -
\frac{p_{\mu}p_\nu}{p^2}\Bigr)
 + {\cal O}(1/d).
\label{A}
\ee
Stress once again that, unlike the partial solutions of 
equations (\ref{D})-(\ref{Dmunu}), formulae 
(\ref{ghost}) and (\ref{A}) define the physical
propagators of particles in the physical 
Poincar{\' e}-invariant vacuum.

As is well known, (see, for instance, \cite{Zin}) 
in  the lattice theories the $1/d$-expansion is, 
in essence, the mean-field expansion.
Probably, as in this instance, the  vacuum of 
Abelian configurations is a peculiar  mean-field
approximation to the true physical vacuum.

In the conclusion of this section let us touch on 
the  cluster properties. For the scheme 
based on an approximation of the physical vacuum by 
the superposition of partial modes the cluster 
decomposition principle is a nontrivial property 
(see, for instance, \cite{Wei}) and requires  
checking at each stage of calculations. We can 
state that this principle is satisfied for our 
model of physical vacuum at least in the leading 
order of $1/d$-expansion. 
So, for instance, from eq.(\ref{F}) we get for the 
second-step four-point function
 ${\cal F}_{\mu\nu\sigma\rho}$ of the gauge field in 
 the leading order in $1/d$:
$
{\cal F}_{\mu\nu\sigma\rho} = 
{\bf D}_{\mu\nu}\otimes{\bf D}_{\sigma\rho} + 
\{\nu\leftrightarrow\sigma\} + \{\nu\leftrightarrow\rho\},
$
which is the usual disconnected part of the four-point 
function, in correspondence with  the cluster
decomposition principle.

\section{Conclusion}

In this paper it is found that for 
non-Abelian $SU(2)$-theories with the physical 
vacuum, representing  a superposition of 
Abelian partial   modes, the gauge field propagator 
in the leading approximation of $1/d$-expansion is 
\be
D^{ab}_{\mu\nu}(p) = \Bigl\{-\frac{1}{p^2}n^an^b + 
\frac{1}{\mu^2 - p^2}(\delta^{ab} - 
n^an^b)\Bigr\} \Bigl(g_{\mu\nu} -
\frac{p_{\mu}p_\nu}{p^2}\Bigr),
\label{Propagator}
\ee
where $\mu^2=-g^2{\cal V}^2$.

If ${\cal V}^2>0$  the spectrum contains tachyons 
which is a sign of instability of this state \cite{Mir}.

At ${\cal V}^2<0$ this propagator corresponds to the 
mass spectrum of $SU(2)$ Georgi-Glashow model 
\cite{GG}, and in this case the considered mechanism 
is a dynamical realization of the gauge sector for 
this model.
As is well known, this model cannot be incorporated in the
Standard Model phenomenology. From the   viewpoint 
of our construction it means that the real physical 
vacuum of the Standard Model has a more complicated 
structure, and for its description it is necessary 
to take into account a wider (or other) class of 
solutions of the characteristic equations --- vacuum 
constituents of fields.
The ensemble of these solutions is highly extensive 
and various, and this variety allows one to hope for a 
possibility of the dynamical  description of the mass 
generation  in the Standard Model on the base of 
principles involved. Phenomenological consequences 
of the dynamical mass generation  in Standard Model 
(see, for instance, \cite{Arb}, \cite{Hun}) lead to 
the interesting physical results, and  further  
studying of the dynamical mass generation mechanism 
 seems  quite actual.
\section*{Acknowledgments}
The author is grateful to A.I.~Alekseev, B.A.~Arbuzov, 
P.A.~Saponov
and L.D.~Soloviev for useful comments.
 The work is supported in part by RFBR, grant No.98-02-16690.

\end{document}